\documentclass[aps,prl,reprint,twocolumn,preprintnumbers,floatfix,nofootinbib]{revtex4-1}

\usepackage{graphicx}
\usepackage{epstopdf}
\usepackage{mathrsfs}
\usepackage{amssymb}
\usepackage{verbatim}
\usepackage{color}
\usepackage[dvipsnames]{xcolor}
\usepackage{multirow}
\usepackage{amsmath}
\usepackage{subfig}
 \usepackage{slashed} 
\usepackage{float}
\usepackage[normalem]{ulem}
\usepackage{sidecap}
\usepackage{hyperref}
\usepackage{cancel}
\usepackage{enumerate}
\hypersetup{pdfstartview=FitV,colorlinks=true,linkcolor=blue,citecolor=red,filecolor=black,urlcolor=blue}

\newcommand{\beq}{\begin{equation}}
\newcommand{\eeq}{\end{equation}}
\newcommand{\bea}{\begin{eqnarray}}
\newcommand{\eea}{\end{eqnarray}}

\def\a{\alpha}
\def\b{\beta}
\def\g{\gamma}

\def\to{\rightarrow}

\def\del{\partial}

\usepackage[parfill]{parskip}
\begin{document}\sloppy 

\preprint{LPT--Orsay 19-01}
\preprint{UMN--TH--3812/19, FTPI--MINN--19/03}

\vspace*{1mm}

\title{Radiative Production of Non-thermal Dark Matter}
\author{Kunio Kaneta$^{a}$}
\email{kkaneta@umn.edu}
\author{Yann Mambrini$^{b}$}
\email{yann.mambrini@th.u-psud.fr}
\author{Keith A. Olive$^{a}$}
\email{olive@umn.edu}
\vspace{0.5cm}
\affiliation{$^a$William I. Fine Theoretical Physics Institute, School of
 Physics and Astronomy, University of Minnesota, Minneapolis, MN 55455,
 USA}
\affiliation{${}^b $ Laboratoire de Physique Th\'eorique (UMR8627), CNRS, Univ.~Paris-Sud, Universit\'e Paris-Saclay, 91405 Orsay, France}

\date{\today}

\begin{abstract} 
We compare dark matter production from the thermal bath in the early universe with its direct production through the decay of the inflaton.
We show that even if dark matter does not possess a direct coupling with the inflaton, Standard Model loop processes may be sufficient to generate the correct relic abundance.
\end{abstract}

\maketitle

\setcounter{equation}{0}

\section{I. Introduction}

Despite indirect but clear evidence \cite{planck} of the presence of large amount of dark matter in our Universe, its nature still remains elusive. The absence of any signal in direct detection experiments XENON \cite{XENON}, LUX \cite {LUX} and PandaX\cite{PANDAX} question the weakly coupled dark matter paradigm. Simple extensions of the Standard Model such as the Higgs-portal \cite{hp,Higgsportal}, $Z$-portal \cite{Zportal}, or even $Z'$-portal \cite{Zpportal} as well as more complex extensions such the minimal supersymmetric standard model \cite{mssm,Go1983,ehnos} have large part of their parameter space (if not all) excluded when combining direct, indirect and accelerator searches (for a review on WIMP searches 
and models, see \cite{Arcadi:2017kky}). 
As a consequence, it is useful to look for different scenarios,
including those with ultra-weak couplings \cite{fimp} (see
\cite{Bernal:2017kxu} for a review), or the possibility that dark matter production occurred at very early stages of reheating after inflation
as in the case of gravitino production \cite{nos,ehnos,kl}.

In this context, it has been confirmed that dark matter production is naturally feasible in a wide variety of models such as SO(10) grand unification \cite{Mambrini:2013iaa}, anomaly free U(1)' models \cite{Bhattacharyya:2018evo}, spin-2 portal \cite{Bernal:2018qlk}
high scale supergravity \cite{Benakli:2017whb,grav2,grav3,gravitino,highsc}
or moduli portal \cite{Chowdhury:2018tzw}. In all these models, it has been shown that effects of non-instantaneous reheating \cite{Giudice:2000ex,Garcia:2017tuj} and non-instantaneous thermalization \cite{Garcia:2018wtq} are non-negligible.

As in the case of gravitino production during inflationary reheating, in many models, dark matter is produced from the annihilation of thermal Standard Model particles, and one neglects the direct production of dark matter $\chi$ from the decay of the inflaton, $\phi$. It was shown \cite{egnop,grav2,grav3} that the branching ratio $BR(\phi \rightarrow \chi \chi)$, is constrained to be very small (below $10^{-8}$ for a 100 GeV dark matter and a reheating temperature $T_{RH} \gtrsim 10^{10}$ GeV). In this paper, we compare the dark matter production rate from both the thermal bath and direct decay. We show that if dark matter is produced from the thermal bath, independent of the tree level branching ratio of inflaton decay to dark matter,  one cannot ignore the radiative decay of the inflaton into dark matter. We further show that the radiative decay may well dominate
the production rate thus providing the main source for relic abundance in the Universe. As a particular example, we also consider the radiative contribution for gravitino production in models of high scale supersymmetry.

The paper is organized as follows. We first compute the dark matter abundance is section II, then apply it to a generic microscopic model in section III before looking to consequences in supersymmetric scenarios in section IV. We conclude in section V.

\section{II. Dark matter production}

\subsection{Generalities}

We presume that the dark matter is not produced in thermal equilibrium during reheating and as a consequence,  the dark matter abundance, $n_{\chi}$, in the early universe is much lower than the thermal density
$n_\gamma$. If we further assume that the dark matter annihilation process,  $n_\chi~n_\chi \rightarrow n_{\gamma} ~n_{\gamma}$, is also out of equilibrium, since $n_\chi^2 \ll n_\gamma^2$, we can write the Boltzmann equation as
\beq
\frac{dn_\chi}{dt} + 3  H n_\chi = R(T)
\label{Eq:boltzmannreheating}
\eeq
with 
\beq
R(T) = \int f_1 f_2 \frac{E_1 E_2 dE_1 dE_2 ~d\cos \theta_{12}}{1024 \pi^6} \int |{\cal M}_{fi}|^2 d \Omega_{13}, 
\label{Eq:rate}
\eeq
for a process 1 2 $\rightarrow$ 3 4 with 1 and 2 
representing standard model particles in the thermal bath and 3, 4 representing the dark matter
candidate, with $f_1$ and $f_2$ being the thermal distribution functions of the incoming particles 1 and 2 and $d\Omega_{13}$ is the solid
angle between the particle 1 and 3 in the center of mass frame.

To compute the relic abundance, the strategy is straightforward. To solve Eq.(\ref{Eq:boltzmannreheating}), one needs to know the relation between $T$ and the cosmological scale parameter, $a$. We can then use 
the Hubble parameter $H = \frac{\dot a}{a}$ 
to express $t$ as function of $T$ before integrating Eq.~(\ref{Eq:boltzmannreheating}). The dependence of $T$ on $a$ is obvious in a pure radiation dominated or a pure
matter dominated era
($T \propto a^{-1}$) due to entropy conservation.  However, during reheating, 
the temperature grows from effectively 0 (when the Universe is still dominated by inflaton oscillations) to a maximum temperature, $T_{\rm max}$, \cite{Garcia:2017tuj,Giudice:2000ex}. Subsequently, 
the temperature decreases to $T_{\rm RH}$ when
the inflaton is (nearly) fully decayed. 
The maximum temperature is determined by the 
inflaton decay width, $\Gamma_\phi$ and is approximated by $T_{\rm max} \simeq 0.5 (M_\phi/\Gamma_\phi)^{1/4} T_{\rm RH}$, where $M_\phi$ is the inflaton mass.
During this period,
the density of Universe is a mixture between the inflaton and the growing radiation density. It has been shown in \cite{Garcia:2017tuj,Giudice:2000ex} that solving the set of conservation of energy conditions for the inflaton density $\rho_\phi$, the radiation density $\rho_R$ combined with the Friedmann equation :
\bea
&&
\frac{d \rho_\phi}{dt} + 3 H \rho_\phi = - \Gamma_\phi \rho_\phi
\nonumber
\\
&&
\frac{d \rho_R}{dt} + 4 H \rho_R=  \Gamma_\phi \rho_\phi .
\label{Eq:reheatingboltzmann}
\\
&&
H^2 = \frac{\rho_\phi}{3 M_P^2} + \frac{\rho_R}{3 M_P^2},
\nonumber
\eea
where $\Gamma_\phi$ is the width of the inflaton\footnote{A more precise computation should distinguish $\Gamma_\phi$ into the part contributing to the thermal bath $\Gamma^\gamma_\phi$ 
and the part contributing to the dark sector $\Gamma^\chi_\phi$. However, we will always consider (and justify) $\Gamma_\phi^\chi \ll \Gamma^\gamma_\phi$ throughout our study.}, 
and $M_P$ the reduced Planck mass\footnote{$M_P = \frac{1}{\sqrt{8 \pi}}M_{Pl}= 2.4\times 10^{18}$ GeV.}
one obtains
\beq
T \propto a^{-3/8} ,
\eeq
as $T$ decreases from $T_{\rm max}$ to $T_{\rm RH}$. 
Defining the comoving yield $Y$
\beq
Y=\frac{n}{T^8} \, ,
\eeq
the Eq.~(\ref{Eq:boltzmannreheating}) becomes
\beq
\frac{dY}{dT} = - \frac{8}{3} \frac{R(T)}{H T^9} \, .
\label{Eq:dydt1}
\eeq

If we further make the approximation that the inflaton dominates the total energy density between $T_{\rm max}$ and $T_{\rm RH}$, we can express the Hubble parameter
in terms of $\rho_\phi$ which falls as $a^{-3} \sim T^8$
with a constant of proportionality depending on the inflaton decay rate,
\beq
H(T) \approx \frac56 \frac{\alpha}{\Gamma_\phi M_P^2} T^4 \, .
\label{h1}
\eeq
In Eq. (\ref{h1}),  $\alpha =\frac{g(T) \pi^2}{30}$, where  $g(T)$ counts the relativistic degrees of freedom of the thermal bath at the temperature $T$ ($g_*=106.75$ in the Standard Model) so that $\rho_R = \alpha T^4$.
In this case, Eq. (\ref{Eq:dydt1}) becomes
\beq
\frac{dY}{dT} =
-\frac{24 c}{5 \sqrt{3 \alpha}} \frac{R(T) M_P T_{RH}^2}{T^{13}}
\label{Eq:dydt}
\eeq
where we have taken 
\beq
\frac32 c H(T_{\rm RH}) = \frac32 c \sqrt{\frac{\alpha}{3}} \frac{T_{\rm RH}^2}{M_P} = \Gamma_\phi
\label{trh}
\eeq
to define the reheating temperature, where
 $c \approx 1.2$ is a constant obtained from a numerical integration \cite{ps2,egnop}. 
The particular choice of a dark matter candidate
and its interactions will determine $R(T)$ and allow
for the integration of Eq. (\ref{Eq:dydt}).

 \subsection{Dark matter production from the thermal bath}

 It has been shown in \cite{Mambrini:2013iaa} that dark matter can be produced in the very early stages of the Universe, even if it is not directly coupled to the Standard Model, through the exchange of a massive mediator. 
 Indeed, thermal gravitino production \cite{nos,ehnos,kl}
 was an early example of this type of process.
 In \cite{Bhattacharyya:2018evo},  \cite{Bernal:2018qlk} \cite{gravitino,grav2,grav3,highsc} and \cite{Chowdhury:2018tzw} this type of dark matter production has been extended to Chern-Simons type couplings, spin-2 mediators, supergravity,
and moduli--portal scenarios respectively taking into account the effects of non instantaneous reheating we discussed above \cite{Garcia:2017tuj, Garcia:2018wtq}. 
 
We can distinguish between the annihilation production processes and the decay rate by defining 
\beq
R(T) = R_{annihilation}(T) + R_{decay}(T) \, .
\eeq
We parametrize the production rate 
\beq
R_{annihilation}(T) = \frac{T^{n+6}}{\Lambda^{n+2}},
\label{Eq:roft}
\eeq

where $\Lambda$ is  some beyond the Standard Model mass scale\footnote{This parametrization corresponds to a cross section $\sigma \propto \frac{T^n}{\Lambda^{n+2}} $.}. 
If one looks at specific models, $n=2$ could correspond to the exchange of a massive particle with mass $\Lambda > T_{\rm max}$ or two non-renormalizable mass suppressed couplings. $n=6$ appears in processes invoking 
 two mass suppression couplings and the exchange of a massive particle, which is the case in high scale supergravity \cite{gravitino,grav2,grav3,highsc} or moduli--portal scenarios \cite{Chowdhury:2018tzw}.

 Inserting Eq.~(\ref{Eq:roft}) in Eq.~(\ref{Eq:dydt}) we obtain, after integration, from $T_{\rm max}$ down to $T_{\rm RH}$,
 for the relic abundance at $T=T_{RH}$ 
 \footnote{These expressions agree (using $c=1$) with the more exact derivation in \cite{Garcia:2017tuj} with the substitution of
 $\Lambda^{n+2} = (\pi/\lambda) (\pi^2/g_\chi \zeta(3))^2 M^{n+2}$ where $g_\chi$ accounts for the internal number of degrees of freedom of the dark matter $\chi$. We note that the cross section was assumed to be $\sigma v = \lambda T^n/\pi M^{n+2}$
 yielding a rate $R = n_\chi^2 \sigma v$, with $n_\chi = g_\chi \zeta(3) T^3/\pi^2$.} :
\bea
&&
\mathrm{n<6:}~~~n(T_{RH}) = \frac{24c}{5} \frac{T_{RH}^{n+4} M_P}{(6-n) \sqrt{3 \alpha}\Lambda^{n+2}}
\nonumber
\\
&&
\mathrm{n=6:}~~~ n(T_{RH}) =  \frac{24c}{5 \sqrt{3 \alpha}} \frac{M_P T_{RH}^{10}}{\Lambda^8} \ln\left( \frac{T_{\mathrm{max}}}{T_{RH}} \right)
\nonumber
\\
&&
\mathrm{n>6:}~~~ n(T_{RH}) = \frac{24c}{5} \frac{T_{\mathrm{max}}^{n-6} M_P T_{RH}^{10}}{(n-6) \sqrt{3 \alpha}\Lambda^{n+2}}
\label{Eq:ntmax}
\eea
from which we can deduce the present relic abundance at the temperature $T_0$
\bea
&&
\Omega  = \frac{\rho}{\rho_c} \approx \frac{n(T_0) \times m_\chi}{10^{-5}h^2 ~ \mathrm{GeV cm^{-3}}}
\label{Eq:omegageneral}
\\
&&
\Rightarrow
\Omega h^2 \approx 10^5~  \frac{n(T_{RH})}{\mathrm{cm^{-3}}} \left( \frac{g(T_0)}{g(T_{RH})}\right) \left(\frac{T_0}{T_{RH}} \right)^3 \left(\frac{m_\chi}{\mathrm{1~GeV}}  \right)
\nonumber
\\
&&
\simeq 5.9 \times 10^6 \left[ \frac{n(T_{RH})}{T_{RH}^3} \right]   \left( \frac{m_\chi}{{\mathrm{1~GeV}}}\right)
\nonumber
\eea

\noindent
with $g(T)$ accounting for the relativistic degrees of freedom, where we have considered only Standard Model 
degrees of freedom ($g(T_{RH}) = 106.75$ and $g(T_0)=3.91$).

 \subsection{Dark matter production from inflaton decay}
 
We can also compute the dark matter density produced directly from the decay of the inflaton. Indeed, in Eq. (\ref{Eq:ntmax}),
 we neglected any direct couplings of the dark matter to the inflaton.
 But we can easily compute the relic abundance
obtained if we allow a branching fraction $B_R$ of the inflaton decay into dark matter. Here $B_R$ is defined as the number of dark matter quanta
produced per inflaton decay.

We will assume non-instantaneous thermalization and solve the Boltzmann equation Eq.~(\ref{Eq:dydt}) defining the production
rate (number of dark matter particles produced per unit of time and volume) as
\beq
R_{decay}(T)=  \Gamma_\phi \frac{\rho_\phi}{M_\phi} B_R =  \frac{25}{12} \frac{\alpha^2~T^8}{\Gamma_\phi M_P^2 M_\phi} B_R \, .
\label{Eq:ratedecay}
\eeq
Plugging the rate Eq.~(\ref{Eq:ratedecay}) into Eq.~(\ref{Eq:dydt}), we obtain 
\bea
\frac{dY}{dT} = - \frac{20}{3} \frac{\alpha}{M_\phi T^5} B_R \, ,
\\
\nonumber
\eea
which gives after integration between
$T_{\rm max}$ and $T_{\rm RH}$,
\bea
&&
Y(T_{RH}) \simeq \frac{5}{3} \frac{\alpha}{M_\phi T_{RH}^4} B_R =  \frac{g(T_{RH}) \pi^2 }{18~M_\phi T_{RH}^4} B_R
\nonumber
\\
&&
~\Rightarrow ~~  n(T_{RH})= \frac{g(T_{RH}) \pi^2 }{18~M_\phi} T_{RH}^4 B_R \, .
\label{Eq:ntrh}
\\
\nonumber
\eea
Interestingly, this approximate result is remarkably close to the 
exact integration done in \cite{egnop} where the coefficient of $\alpha T_{\rm RH}^4 B_R/M_\phi$ in $n$ is $(8/3) (c/3)^{1/2}$ which is almost exactly
5/3 when $c = 1.2$ \footnote{Notice that if
we considered instantaneous reheating, one would have
$n(T_{RH})=
B_R \times \frac{\rho_\phi}{M_\phi}  = B_R \times \frac{g(T_{RH})\pi^2}{30 M_\phi} T_{RH}^4$.}.

We can then combine Eq.~(\ref{Eq:omegageneral}) with Eq.~(\ref{Eq:ntrh}) to obtain
\bea
&&
\Omega h^2|_{\mathrm{decay}} = \left( \frac{B_R}{2.9 \times 10^{-9}} \right)
\left( \frac{T_{RH}}{M_\phi} \right) \left( \frac{m}{1~ \mathrm{GeV}} \right)
\label{Eq:omegadecay}
\\
&&
\simeq 0.1 \left( \frac{B_R}{8.7 \times 10^{-7}} \right) \left(\frac{3 \times 10^{13} \mathrm{GeV}}{M_\phi} \right) 
 \left( \frac{T_{RH}}{10^{10} \mathrm{GeV}}\right) 
\left( \frac{m}{{\mathrm{1~GeV}}}\right)
\nonumber
\eea

As we indicated earlier, we observe that 
one needs a very low branching ratio to avoid the overabundance of the dark matter. 
The total relic abundance is then obtained by adding the annihilation and decay contributions. Combining Eq.~(\ref{Eq:ntmax}) and
Eq.~(\ref{Eq:omegadecay}), we obtain

\beq
\Omega h^2|_{\mathrm{tot}} \simeq \Omega h^2 |_{annihilation} + \Omega h^2|_{\mathrm{decay}}
\nonumber
\eeq

\noindent
or
\begin{widetext}
\bea
&&
\Omega h^2|_{n<6} \simeq 0.1
 \left[
 \frac{2.8 \times 10^8 c}{(6-n)} ~ \frac{T_{RH}^{n+1} M_{P}}{\sqrt{3 \alpha}~ \Lambda^{n+2}} 
 + \frac{B_R}{2.9 \times 10^{-10}}   \times \frac{T_{RH}}{M_\phi}
 \right] 
 \left( \frac{m}{{\mathrm{1~GeV}}}\right)
\nonumber
\\
&&
\Omega h^2|_{n=6} \simeq 0.1
 \left[
  2.8 \times 10^8 c  ~ \frac{T_{RH}^7 M_P}{\sqrt{3 \alpha}~\Lambda^8}
  \ln \left(\frac{T_{max}}{T_{RH}} \right)
  + \frac{B_R}{2.9 \times 10^{-10}}   \times \frac{T_{RH}}{M_\phi}
  \right]
 \left( \frac{m}{1~ \mathrm{GeV}}\right)
\nonumber
\\
&&
\Omega h^2|_{n>6} \simeq 0.1
 \left[
 \frac{2.8 \times 10^8 c}{(n-6)} ~ \frac{T_{RH}^{7} T_{max}^{n-6} M_{P}}{\sqrt{3 \alpha}~ \Lambda^{n+2}} 
 + \frac{B_R}{2.9 \times 10^{-10}}   \times \frac{T_{RH}}{M_\phi}
\right] \left( \frac{m}{{\mathrm{1~GeV}}}\right)
\nonumber
\\
&&
\label{Eq:omegatot}
\eea
\end{widetext}
which gives, for $n=6$ for example
\begin{widetext}
\beq
\Omega h^2|_{n=6}^{tot}\simeq 0.1 \left[ \left( \frac{T_{\rm RH}}{10^{11}{\rm GeV}} \right)^7 \left( \frac{7.3 \times 10^{12}{\rm GeV}}{\Lambda} \right)^8 
\ln \left(\frac{T_{\rm max}}{T_{\rm RH}} \right)
+ \left( \frac{B_R}{8.7 \times 10^{-8}} \right) \left( \frac{3 \times 10^{13}{\rm GeV}}{M_\phi}\right) \left( \frac{T_{\rm RH}}{10^{11} {\rm GeV}} \right)  \right] 
\left( \frac{m}{1~\mathrm{GeV}}\right) .
\eeq
\end{widetext}

\begin{figure}[h!]
\centering
\includegraphics[scale=0.6]{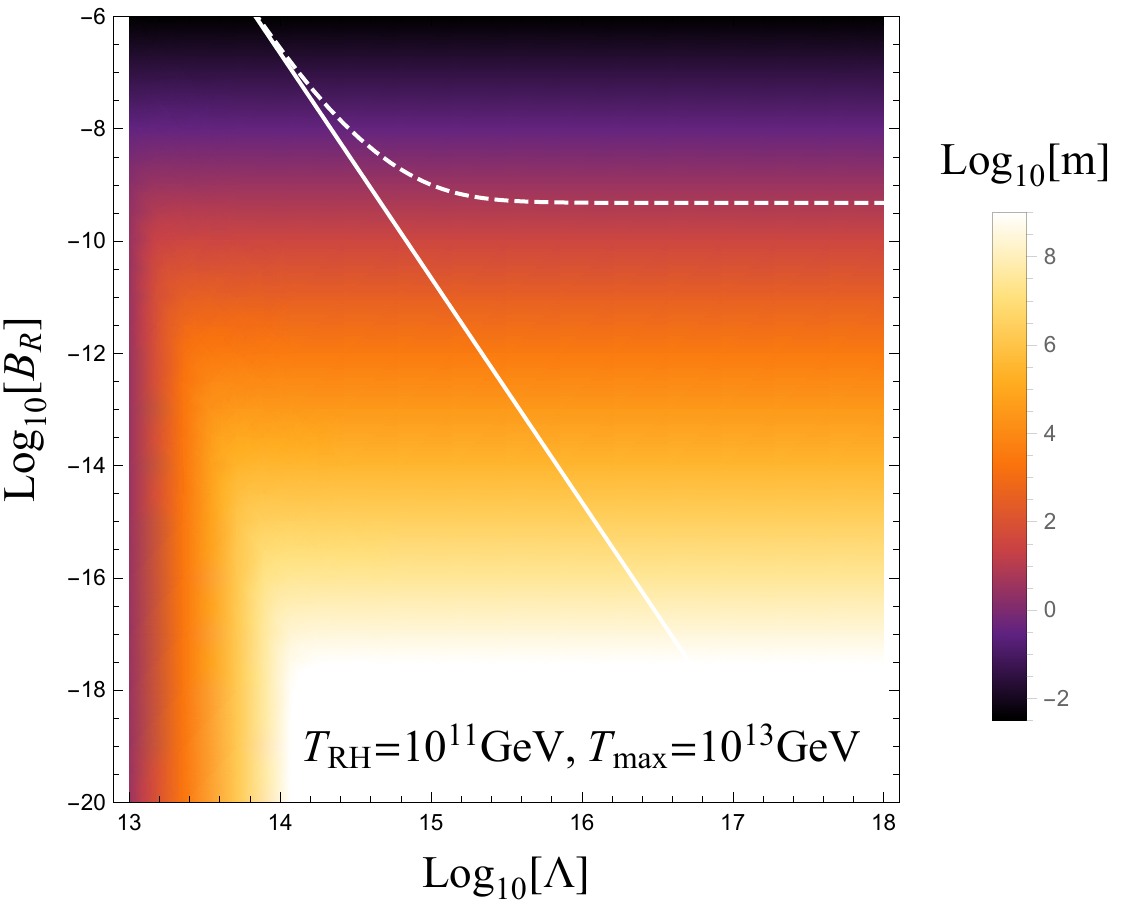}
\caption{\em \small Parameter space in the plane ($\Lambda$, $B_R$) of points respecting PLANCK constraints, 
with the corresponding dark matter mass $m$
for the case with $n=6$.  The lines correspond to the branching ratio determined in our microscopic approach for $y_\chi=0$ (white solid line) and $y_\chi = 10^{-4} y_f$ (white dashed line). See the text and Eq. (\ref{Eq:brmicro}) for details. }
\label{Fig:brmicro}
\end{figure}

We show in Fig. \ref{Fig:brmicro} the result on the scan of the parameter space (in $B_R$, $\Lambda$, and the dark matter mass, $m$) in the case $n=6$ using the relic density constraint 
$\Omega h^2 \simeq 0.11$, for $T_{RH} =10^{11}$ GeV and $T_{max} = 10^{13}$ GeV. For each value of $\Lambda, B_R$, the mass needed to obtain the correct relic density is color coded by the scale at the right of the figure. We notice 
that the range of dark matter masses allowed is very large (from the MeV to the EeV scale). This is a direct consequence of the strong power dependence
of the relic abundance on the scale $\Lambda$. We also remark that a branching ratio of $\mathcal{O}(1)$ is possible for dark
matter masses of order 100 keV,  whereas EeV scale dark matter requires very tiny branching fraction of the order of 
$B_R \simeq 10^{-16}$ to avoid the over-closure of the Universe. Note also, that over most of the parameter space with large values of $\Lambda$, the thermal production from annihilations contributes negligibly to
the relic density. 

\section{III. Generic microscopic models}

We next apply the generic analysis we did using an effective field theory formulation. 
We first consider the
effective Lagrangian\footnote{Without loss of generalities, we will work with a fermionic dark matter candidate. The extension to a scalar candidate is straightforward.} between the inflaton $\phi$, 
standard model fields, $f$, and the dark matter candidate, $\chi$:
\beq
{\cal L} = y_f \phi \bar f f + y_\chi \phi \bar \chi \chi 
+\frac{1}{\Lambda^2} \bar f f \bar \chi \chi
\label{Eq:lagrangian}
\eeq
This potential can be viewed as an effective interaction between standard model fermions and the dark sector through the exchange of a massive field of mass $M$ $\simeq \Lambda\gg T_{max}$.

\begin{figure}[h!]
\centering
\includegraphics[scale=0.5]{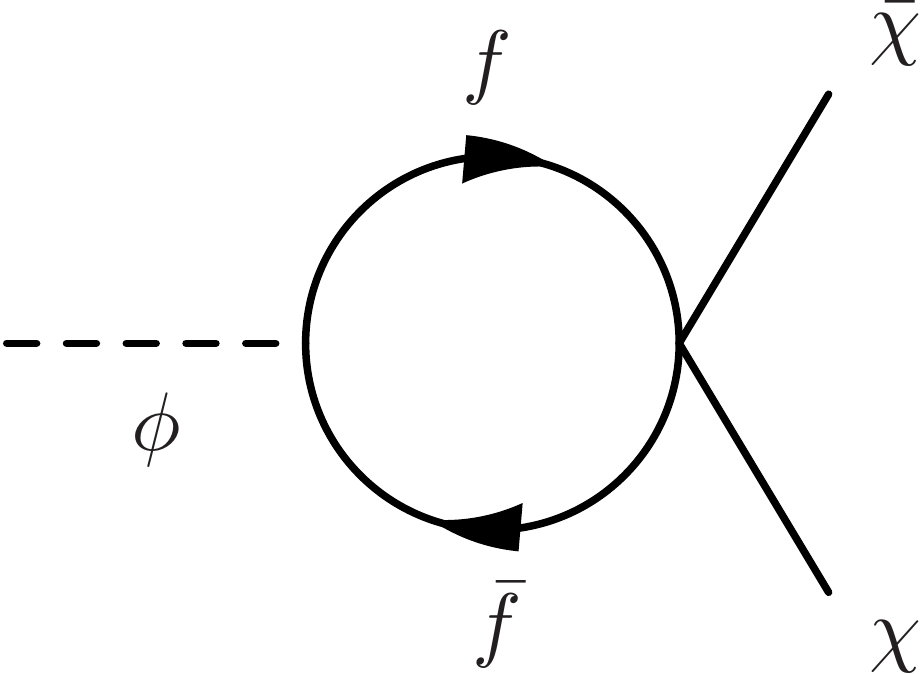}
\caption{\em \small The radiative decay of the inflaton into a pair of dark matter fields.}
\label{Fig:feyn}
\end{figure}

One of our key points is
that there will be direct production of dark matter through inflaton decay, even if the dark matter does not couple
to the inflaton at tree level.
That is, even if $y_\chi = 0$, dark matter will be 
produced radiative via the diagram shown in 
Fig. \ref{Fig:feyn}.  
Indeed, from Eq.~(\ref{Eq:lagrangian})
one can deduce
\beq
\Gamma_\phi^{loop} \simeq N_f\frac{y_f y_\chi}{16\pi^3}\frac{M_\phi^3}{\Lambda^2}+N_f\frac{y_f^2}{128\pi^5}\left(1+\frac{\pi^2}{4}\right)\frac{M_\phi^5}{\Lambda^4}
\eeq
via the loop of the $N_f$ families of the standard model fermions $f$, where the first term is the interference between the tree and loop diagrams, and we have taken the massless limit for $f$.
The other contributions to the inflaton width $\Gamma_\phi$ are given by
\beq
\Gamma_\phi^f= \frac{N_f ~y_f^2}{8 \pi} M_\phi;
~~ \Gamma_\phi^\chi = \frac{y_\chi^2}{8 \pi} M_\phi \, 
\label{Eq:gammas}
\eeq

We can express $B_R$  as function of the microscopic parameters, 
\bea
B_R &=& \frac{\Gamma_\phi^\chi + \Gamma_\phi^{loop}}{\Gamma_\phi^f +\Gamma_\phi^\chi + \Gamma_\phi^{loop}}\nonumber\\
&\simeq&
\frac{1}{N_f} \left(\frac{y_\chi}{y_f}  \right)^2 + \frac{1}{2\pi^2}\frac{y_\chi}{y_f}\left(\frac{M_\phi}{\Lambda}  \right)^2 \nonumber \\ 
&& + \frac{1}{16 \pi^4}
\left(1+\frac{\pi^2}{4}\right)
\left(\frac{M_\phi}{\Lambda}  \right)^4
\label{Eq:brmicro}
\eea
where we assumed that $N_f y_f \gg y_\chi$ and $\Lambda \gg M_\phi$.
These conditions are necessary to avoid the overabundance of dark matter.
Also plotted in Fig. \ref{Fig:brmicro} is the corresponding branching ratio in the plane ($\Lambda$, $B_R$) for two different values of $y_\chi$
($y_\chi=0$ and $y_\chi=10^{-4} y_f$).
First of all, we notice that the lines
lie in the region where the decay of the inflaton is 
responsible of the total amount of dark matter. The thermal production does not contribute in this area. Secondly, the plateau seen by the dashed
curve corresponds to the points where the dark matter is completely produced by the 
tree level decay, uniquely determined by $y_\chi$ and thus 
independent of $\Lambda$. Moreover, even if $y_\chi=10^{-4}y_f$, for low
values of $\Lambda \lesssim 5 \times 10^{14}$ GeV, the loop contribution dominates over the tree-level production.

\section{IV. Supergravity}


We next consider a more concrete supergravity model where the inflaton decays to only a pair of Higgs bosons at tree level, while a pair of gravitinos are produced via the loops of Higgses and Higgsinos.
As a specific application, we would like to 
reconsider the high scale supergravity model described in \cite{grav3}.
The model is based on no-scale supergravity \cite{no-scale}.The inflationary and supersymmetry breaking sector contains
three chiral fields, $T, \phi$ driving inflation \cite{ENO6}, and a Polonyi-like field, $z$ \cite{pol} whose superpotential we take
\beq
W \; = \; \sqrt{3}m_{3/2}(z+\nu)
\eeq
with $\nu$ being a constant.
The Polonyi field is assumed to be twisted and strongly stabilized \cite{dine,ego,dlmmo} so that the K\"ahler potential takes the form
\beq
K \; = \; - 3 \ln \left(T + {\bar T} - \frac{1}{3} \sum_i |\phi_i|^2\right) + |z|^2 - \frac{|z|^4}{\Lambda_z^2}  \,,
\label{Kpol3}
\eeq
where one of the $\phi_i$ is related to inflation, and the rest are matter fields.
Choosing \cite{Cecotti},
\beq\label{tph_w}
W=\sqrt{3} M_T \phi(T-1/2)  \, ,
\eeq
leads to Starobinsky-like inflation \cite{Staro} and is consistent with 
Planck observations \cite{planck} when the inflaton mass, which we now designate as $M_T \simeq 3 \times 10^{13}$ GeV. 

If $\phi$ in (\ref{tph_w}) is fixed so that 
$\langle \phi \rangle = 0$, the inflaton is associated with $T$ and its coupling to Standard Model fields leads to reheating, where the canonically normalized inflaton field $t$ is defined as $T\simeq(1/2)(1+\sqrt{2/3}t)$
about the minimum of $\langle T \rangle = 1/2$.
In a high scale supersymmetry model, all of the superpartners, except the gravitino, are assumed to be more massive than the inflaton. 
In particular, the MSSM $\mu$-parameter is also large, $\mu > M_T$. In this case,
the dominant decay mode for the inflaton
is $t\to H_{u,d} {H^*}^{u,d}$
which ultimately corresponds to a decay of $t \to h h$ where 
$h$ is the SM Higgs boson. The decay rate to two Higgs bosons is \cite{EGNO4}
\beq
\Gamma_{2h} = \frac{\mu^4}{48\pi M_T M_P^2}\equiv \frac{y^2}{8\pi}M_T\,
\eeq
with $y^2\equiv \mu^4/(6m_T^2M_P^2)$.
Then the reheating temperature can be expressed as $T_{\rm RH}\simeq 0.5 (y/2\pi)\sqrt{M_T M_P}$ \cite{grav2,egnop,Pradler:2006hh} and $T_{\rm max} \simeq 0.5 (8\pi/y^2)^{1/4} T_{\rm RH}$.

Gravitinos are produced by the pair annihilation of SM particles, ${\rm SM}~{\rm SM}\to$ two gravitinos, where the resultant gravitino abundance depends on the reheating process as discussed above.
The reaction rate of the gravitino production is given by \cite{Benakli:2017whb}
\bea
R &\simeq& \frac{21.65}{9}\frac{T^{12}}{m_{3/2}^4M_{\rm P}^4}\equiv \frac{T^{12}}{\Lambda^8}
\eea
with $\Lambda^8\equiv 9m_{3/2}^4M_{\rm P}^4/21.65$.
Then, from Eqs. (\ref{Eq:ntmax}) and (\ref{Eq:omegageneral}), we obtain
\bea
\Omega h^2|_{\rm ann}^{\rm inst} &=&
1.9 \times10^{25}\frac{T_{RH}^7}{m_{3/2}^3M_P^4}\ln\left[\frac{T_{\rm max}}{T_{\rm RH}}\right]\nonumber\\
&\simeq&0.12
	\left(
		\frac{0.1~{\rm EeV}}{m_{3/2}}
	\right)^3
	\left(
		\frac{y}{2.3 \times10^{-5}}
	\right)^{7}\nonumber\\
& \times & \left(\frac{M_T}{3\times10^{13}~{\rm GeV}}\right)^{7/2}	\ln\left(1.1/y^{1/2}\right)	,
\label{inst}
\eea
where the instantaneous thermalization of SM particles is implicitly assumed.
On the other hand, when the non-instantaneous thermalization effect in the gravitino production is incorporated, we have
\cite{Garcia:2018wtq}
\bea
\Omega h^2|_{\rm ann}^{\rm non-inst} &=& 0.12
\left(\frac{0.1~{\rm EeV}}{m_{3/2}}\right)^3
\left(\frac{M_T}{3\times10^{13}~{\rm GeV}}\right)^{67/10}\nonumber\\
&&
\times
\left(\frac{y}{3.0\times10^{-7}}\right)^{19/5}
\left(\frac{0.03}{\alpha_3}\right)^{16/5}
\label{noninst}
\eea
where only gluon pair annihilation was assumed.
We evaluate the SU(3)$_C$ gauge coupling $\alpha_3=g_s^2/4\pi$ at $T_{RH}$ by solving renormalization group equations at two-loop level.

In this model, gravitinos are also produced by the tree-level decay of inflaton.
Although the tree level coupling between gravitino and inflaton vanishes when $\langle \phi\rangle=0$, supersymmetry breaking shifts $\langle\phi\rangle$, giving rise to the tree level decay given by\footnote{By comparing to the result of Ref.~\cite{EGNO4}, the given expression is corrected by taking into account higher order terms of vacuum expectation values of $T,~\phi,$ and $z$, and the mixing between $t$ and $z$.}
\beq
\Gamma^{tree}_t = \left(\frac{\Lambda_z}{M_P}\right)^4\frac {81 m_{3/2}^2M_T}{128\pi M_P^2}.
\eeq

\begin{figure}[h!]
\centering
\includegraphics[scale=0.4]{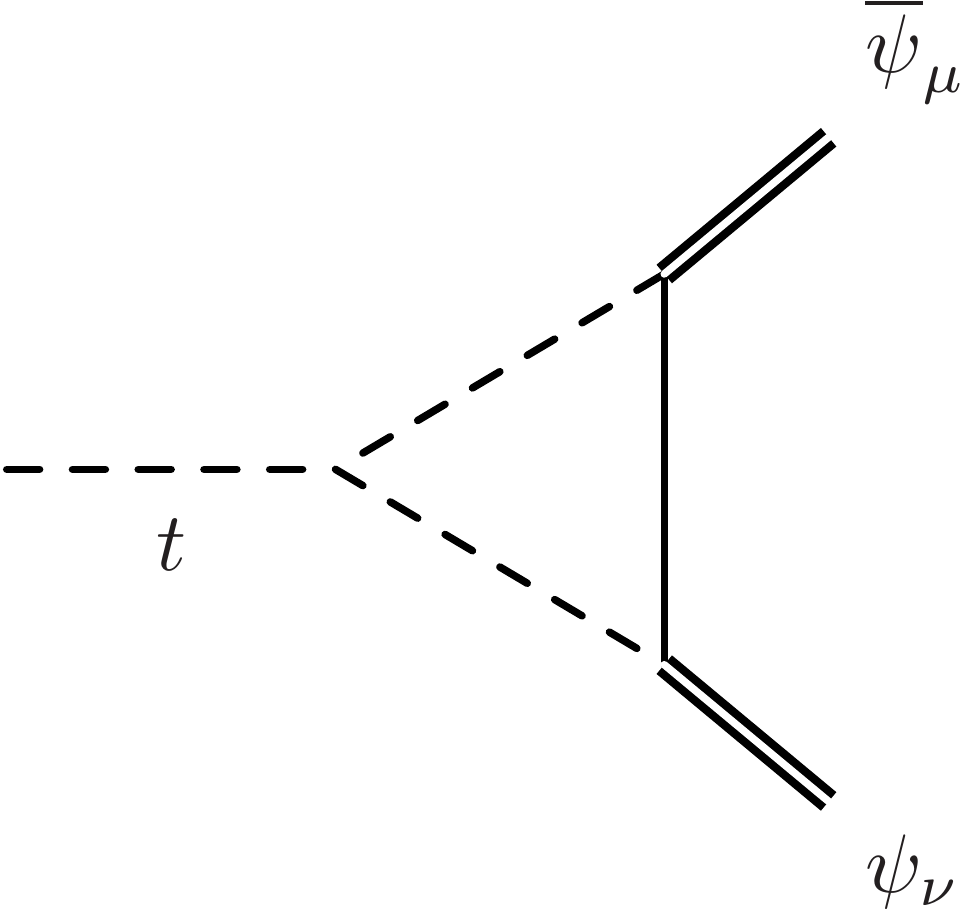}
\includegraphics[scale=0.4]{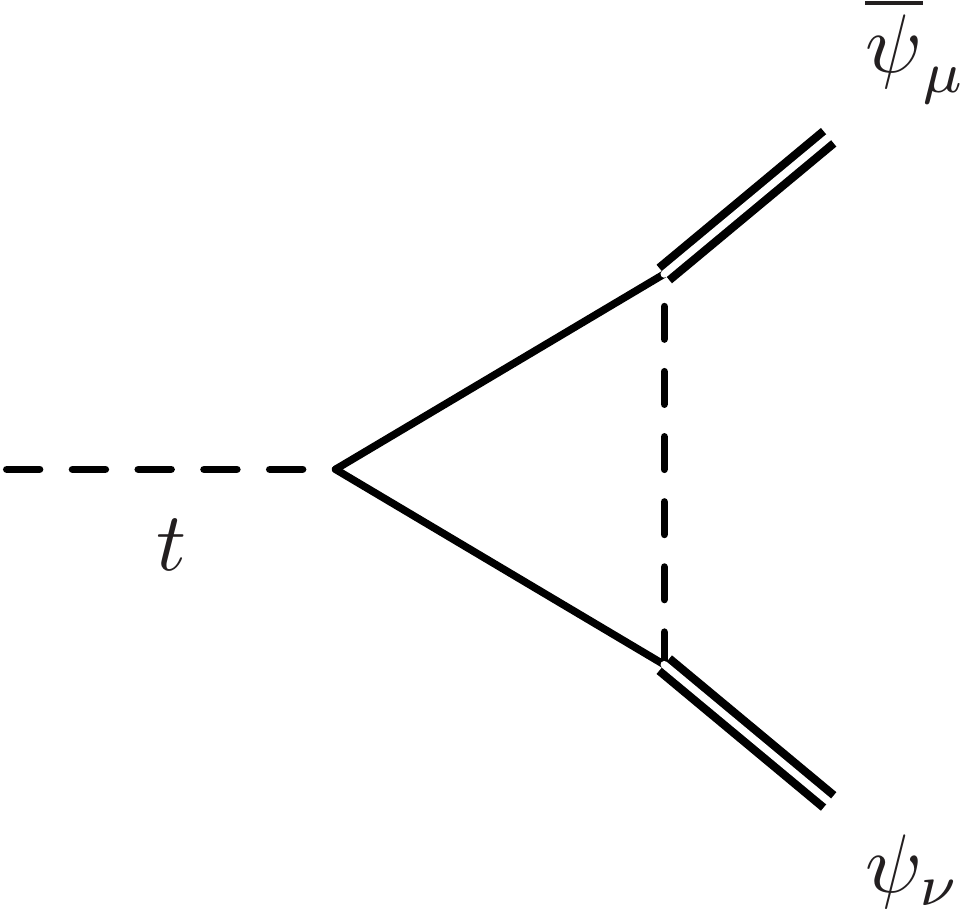}\\ \vspace*{5mm}
\includegraphics[scale=0.4]{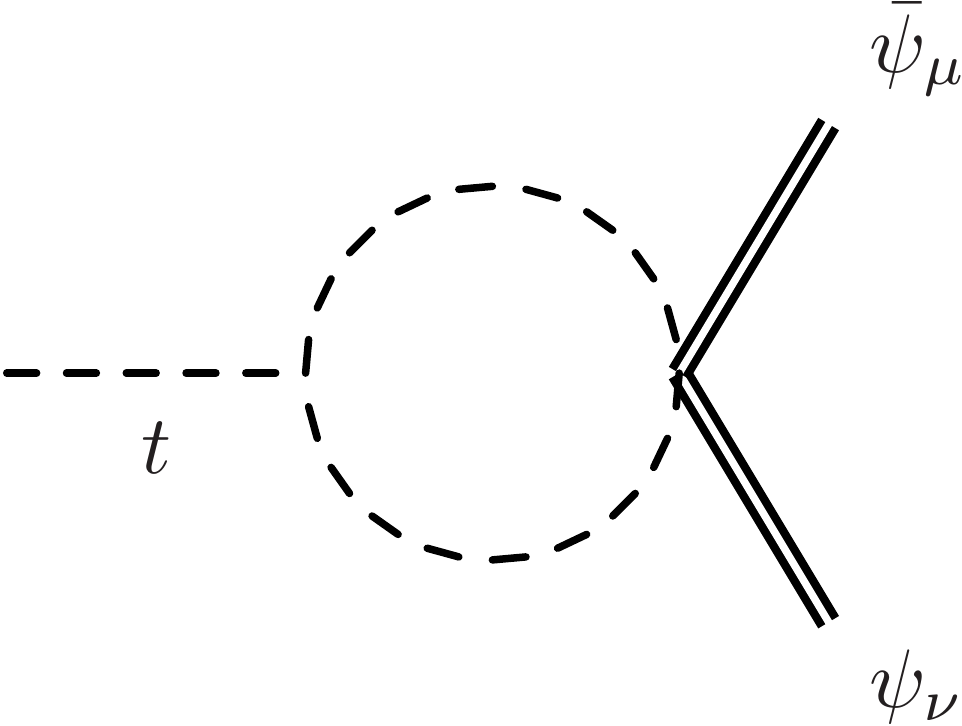}
\includegraphics[scale=0.4]{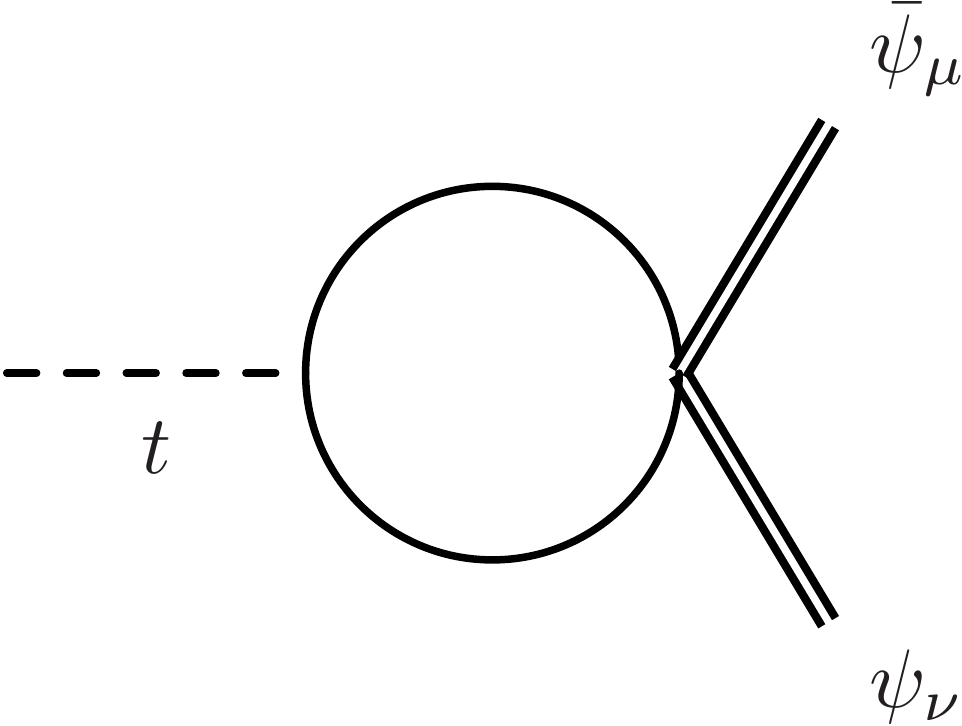}
\caption{\em \small The radiative decay of inflaton into a pair of gravitinos. Dashed (solid) lines in the loop represent the Higgs bosons (Higgsinos). We label those diagrams as A (top left), B (top right), C (bottom left), and D (bottom right).}
\label{Fig:grav}
\end{figure}

The radiative decay of inflaton to a pair of gravitinos is induced by the interactions given by
\beq
{\cal L} \supset C_{tSS}~t(|H_u|^2+|H_d|^2)+C_{tFF}~t(\overline{\widetilde H_u}\widetilde H_d+h.c.),
\eeq
where $C_{tSS}\equiv \sqrt{2/3} \mu^2$ and $C_{tFF}\equiv \mu^2/2\sqrt{6}$ \cite{EGNO4}.\footnote{Note that those interaction terms are obtained by using equations of motion. A more rigorous calculation would not use equations of motion when the interaction terms are relevant for loop diagrams. However, a more rigorous treatment of this particular process is not our main focus, and is expected to have only a small quantitative effect on our result.}
The other relevant terms in the supergravity Lagrangian are given by
\bea
{\cal L} &\supset& -\frac{i}{\sqrt{2}M_P}(\partial_\mu\varphi^*)\overline\psi_\nu\gamma^\mu\gamma^\nu\chi_L+h.c.
\eea
where we denote a chiral multiplet $(\varphi,\chi_L)$.
The relevant diagrams for the radiative decay of inflaton $t$ into a pair of gravitinos are shown in Fig. \ref{Fig:grav}, where the dominant contribution is coming from the upper two diagrams (A and B).
While a detailed discussion of the decay width is given in Appendix~\ref{Sec:appendix}, in the case of $m_{3/2}\ll M_T \ll \mu$ we obtain an approximate expression given by
\beq
\Gamma^{loop} \simeq \frac{2}{3^34^5\pi^5}\left(\frac{1}{4}-\ln\frac{\mu^2}{\mu_{\rm ren}^2}\right)^2\frac{\mu^4M_T^5}{m_{3/2}^2M_P^6},
\label{Eq:LoopWidth}
\eeq
where $\mu_{\rm ren}$ is the renormalization scale, and we take $\mu_{\rm ren}=M_T$ in our analysis.
Then, by using
\bea
B_R &\simeq& \frac{2(\Gamma^{tree}+\Gamma^{loop})}{\Gamma_{2h}} \equiv B_R^{tree}+B_R^{loop},\\
B_R^{tree} &=& \frac{243}{4}\left(\frac{\Lambda_z}{M_P}\right)^4\frac{m_{3/2}^2M_T^2}{\mu^4}\nonumber\\
&\simeq& 5.5\times10^{-12}
	\left(
		\frac{\Lambda_z}{M_P}
	\right)^4
	\left(
		\frac{m_{3/2}}{0.1~{\rm EeV}}
	\right)^2
	\nonumber\\
&& \times
	\left(
		\frac{M_T}{3\times10^{13}~{\rm GeV}}
	\right)^2
	\left(
		\frac{10^{14}~{\rm GeV}}{\mu}
	\right)^4, \label{Btree} \\
B_R^{loop} &\simeq& \frac{1}{144\pi^4}
\left(\frac{1}{4}-\ln\frac{\mu^2}{M_T^2}\right)^2
\frac{M_T^6}{m_{3/2}^2M_P^4}\nonumber\\
&\simeq&
9.8\times10^{-15}
	\left(
		\frac{0.1~{\rm EeV}}{m_{3/2}}
	\right)^2
	\left(
		\frac{M_T}{3\times10^{13}~{\rm GeV}}
	\right)^6\nonumber\\
&& \times
	\left[
		1-8\ln\left(\frac{\mu}{M_T}\right)
	\right]^2,
	\label{Bloop}
\eea
we can evaluate $\Omega h^2|_{\rm decay}$ given in Eq.~(\ref{Eq:omegadecay}).

\begin{figure}[h!]
\centering
\includegraphics[scale=0.5]{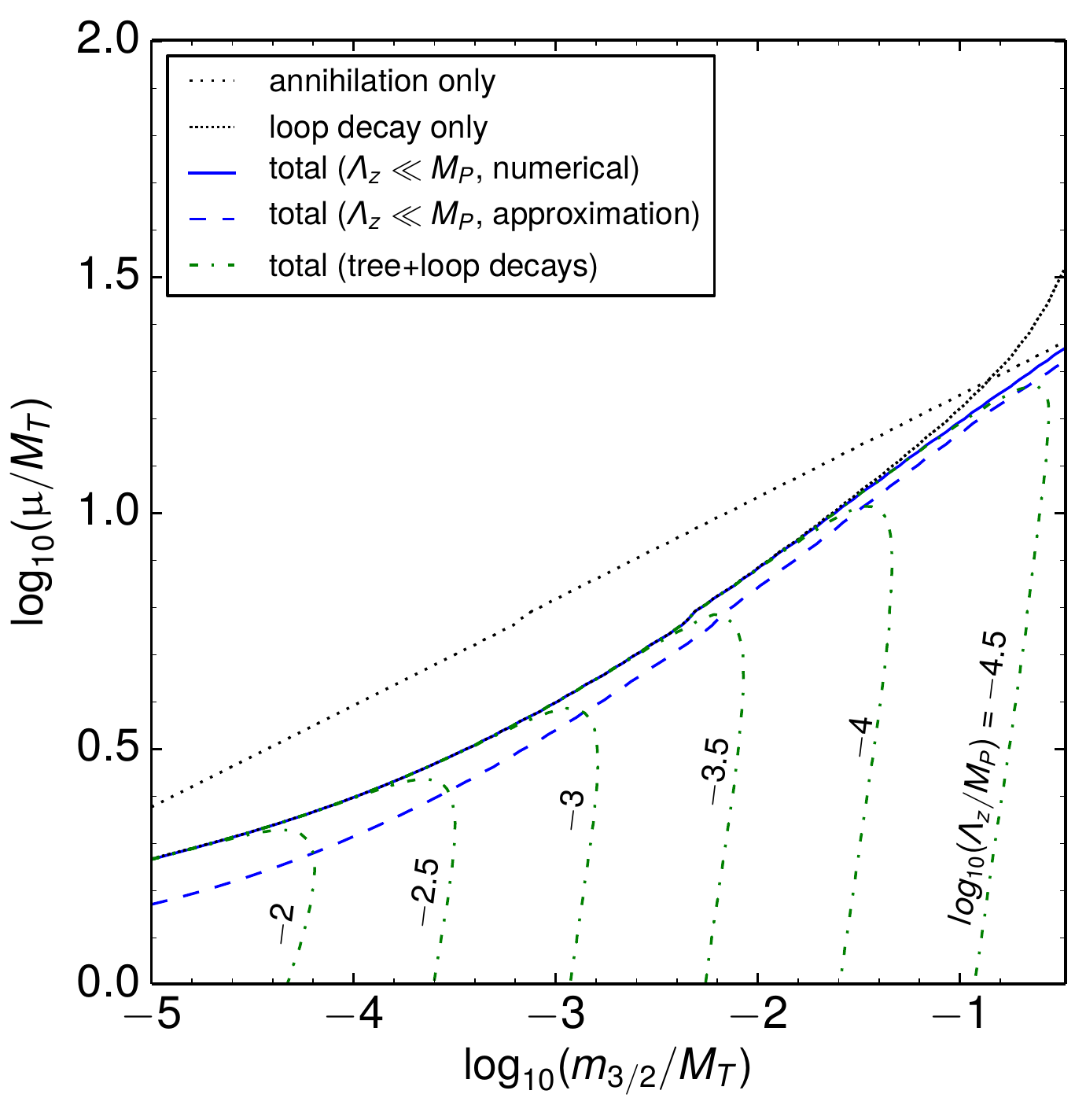}
\includegraphics[scale=0.5]{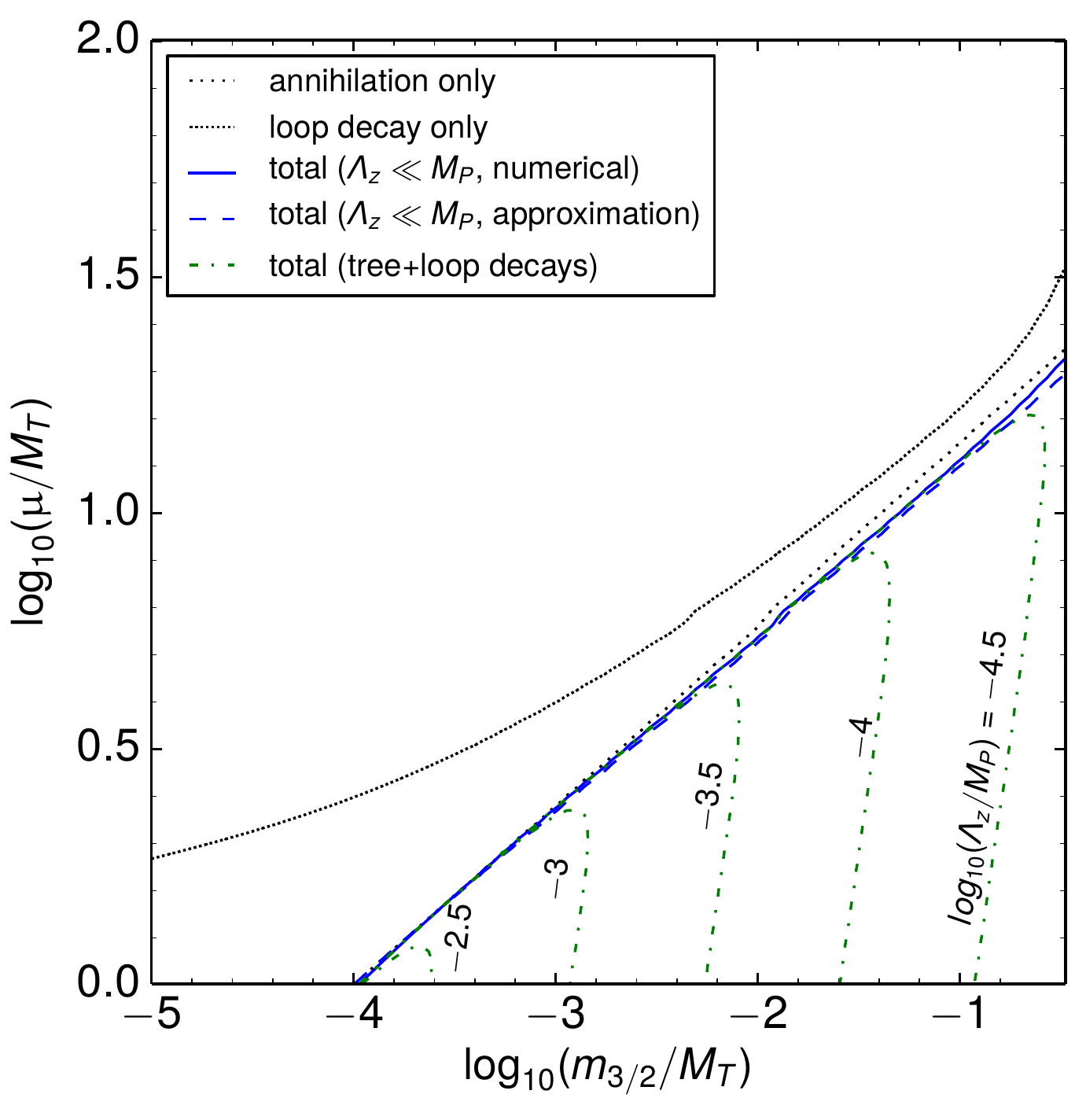}
\caption{\em \small The total number density of gravitino. The top panel assumes the instantaneous thermalization, and the bottom panel takes into account the effect of  non-instantaneous thermalization.}
\label{Fig:Oh2}
\end{figure}

Figure \ref{Fig:Oh2} shows the required relation between the $\mu$-parameter and the gravitino mass when total gravitino abundance given by $\Omega h^2|_{\rm tot}=\Omega h^2|_{\rm ann}^{\rm inst/non-inst}+\Omega h^2|_{\rm decay} = 0.12$, where we take $M_T=3\times10^{13}$ GeV.
The top panel shows the case assuming instantaneous thermalization using Eq. (\ref{inst}), and the bottom panel takes the non-instantaneous thermalization effect into account using Eq. (\ref{noninst}).  
In the both panels, the black dotted line shows the contribution from thermal production alone, using only $\Omega h^2|_{\rm ann}$.
The short spaced dotted line shows the relation with only the loop contribution included. 
As one can see, the contribution from the 1-loop decay 
diagrams dominate over the thermal annihilation processes
when $m_{3/2} \lesssim 10^{-1} M_T$.
The solid blue contour corresponds the sum of the annihilation and loop decay contributions in  $\Omega h^2|_{\rm tot}=0.1197$ \cite{planck} evaluated without the approximation on the Passarino-Veltman functions \cite{Passarino:1978jh}, while the dashed blue line uses the approximate formula given in Eq. (\ref{Eq:LoopWidth}).
To evaluate the Passarino-Veltman functions, we have utilized {\tt LoopTools} \cite{Hahn:1998yk}.
For both blue lines, we take the tree-level decay contribution to be negligibly small by assuming $\Lambda_z\ll M_P$.
The green dot-dashed lines are the total abundance with the tree level decay. The region below the solid blue line is the allowed parameter space.

\section{V. Conclusion}

The origin of the dark matter relic density is unknown.
The thermal production mechanism \cite{hlw}, works quite well for weakly interacting massive particles.
However, this mechanism does not work when
the dark matter candidates are super-weakly interacting
and decouple very early (before inflation) in the history 
of the Universe. Gravitinos are a well known example of
dark matter candidates for which their relic abundance
is not determined by thermal annihilations. 
Instead, these particles are produced during the reheating 
process after inflation. 

Some fraction of the dark matter may also be produced directly from inflaton decay if the inflaton couples to
the dark matter candidate. However, even if the inflaton 
is not directly coupled to the dark matter, we have shown
that direct production necessarily still occurs through
radiative processes as in Fig. \ref{Fig:feyn}. 
We further showed in this work that direct production of dark matter through the radiative decay of the inflaton can in fact dominate the relic abundance in the Universe. 

We highlighted this result with a specific example from
no-scale supergravity. The model makes use of no-scale 
Starobinsky-like inflation, with supersymmetry breaking
through a strongly stabilized Polonyi mechanism. 
We considered the case of high-scale supersymmetry in which
all superpartners except the gravitino are more massive than
the inflaton. Therefore, the thermal content of the universe
is just that of the Standard Model. The dominant decay of the inflaton in this is model is to a pair of Standard Model
Higgs bosons. The branching fraction of direct decays to gravitinos is suppressed (as given in Eq. (\ref{Btree})),
though the degree of suppression depends on the strong stabilization parameter $\Lambda_z$.
However the branching fraction to gravitinos through the loop process shown in Fig. \ref{Fig:grav} and given in 
Eq. (\ref{Bloop}) dominates over the thermal production 
when $m_{3/2} < 10^{-1} M_T \sim 3 \times 10^{12}$ GeV.

\vskip.1in
{\bf Acknowledgments:}
\noindent 
The authors want to thank especially E. Dudas and M. Garcia for very  insightful
discussions. This research has been supported by the (Indo-French)
CEFIPRA/IFCPAR Project No.~5404-2. Support from CNRS LIA-THEP and the
INFRE-HEPNET of CEFIPRA/IFCPAR is also acknowledged.  This work was also supported by the France-US PICS MicroDark.  
Y. Mambrini acknowledges
partial support from the European Union Horizon 2020 research and
innovation programme under the Marie Sklodowska-Curie: RISE
InvisiblesPlus (grant agreement No 690575) and the ITN Elusives (grant
agreement No 674896).
The work of K. Kaneta and K. A. Olive was supported in part by the DOE grant DE-SC0011842 at the University of Minnesota.

\section*{Appendix}
\section{Calculation of the radiative gravitino decay}
\label{Sec:appendix}

The relevant gravitino interactions in the supergravity Lagrangian\footnote{See, for instance, Ref. \cite{Lsugra}} that involves Higgs and Higgsino are given by
\bea
{\cal L}_{\rm 3pt} &=& -\frac{i}{\sqrt2}\del_\mu\varphi^{*}\overline\psi_\nu\gamma^\mu\gamma^\nu\chi_L+\frac{i}{\sqrt2}\del_\mu\varphi\overline\chi_L\gamma^\nu\gamma^\mu\psi_\nu,\label{Eq:L3pt}\\
{\cal L}_{\rm 4pt} &=&
\frac{1}{8}\epsilon^{\mu\nu\rho\sigma}\overline\psi_{\mu}\gamma_\nu\Phi_\rho\psi_{\sigma}-\frac{1}{4}e^{G/2}\overline\psi_{\mu}[\gamma^\mu,\gamma^\nu]\psi_{\nu},\label{Eq:L4pt}\\
{\cal L}_{\rm 4F} &=& -\frac{1}{8}G_{ij^*}
	\overline\psi_{\mu}
	\left[
		i\epsilon^{\mu\nu\rho\sigma}\gamma_\nu+g^{\mu\rho}\gamma^\sigma\gamma^5
	\right]
	\psi_\rho(\overline\chi^j_L\gamma_\sigma\chi^i_L),\nonumber\\
	\label{Eq:L4F}
\eea
where $G\equiv K+\ln|W|^2$, $G_i (G_{ij^*})$ represents derivative of $G$ with respect to $\varphi^i$ ($\varphi^i$ and $\varphi^{*j}$), and $\Phi_\rho\equiv G_i\del_\rho\varphi^i-G_{i^*}\del_\rho\varphi^{*i}$ which vanishes for $i=H_u$ and $H_d$.
We also have $G_{H_u H_u^*}=G_{H_d H_d^*}\simeq 1$.
The second term of Eq. (\ref{Eq:L4pt}) gives Higgs-Higgs-gravitino-gravitino coupling.
In the case of $m_{3/2}\ll \mu$, the dominant terms are induced by
\beq
\langle e^{G/2}\rangle \supset \frac{1}{2}\mu(H_uH_d+H_u^\dagger H_d^\dagger)\, .
\eeq
Thus, we have the four diagrams, shown in Fig. \ref{Fig:grav}, that induce the radiative inflaton decay into a pair of gravitinos.

In calculating the decay amplitudes, the gravitino equation of motion (EOM) may be used to simplify the expressions, which is given as
\begin{equation}
    \epsilon^{\mu\nu\rho\sigma}\gamma^5\gamma_\nu\del_\rho\psi_\sigma+\frac{1}{2}m_{3/2}[\gamma^\mu,\gamma^\nu]\psi_\nu = 0.
\end{equation}
Using the EOM, we also obtain the following relations:
\begin{eqnarray}
	&&(\cancel{p}-m_{3/2})\psi_\mu(p) = 0,\label{Eq:eom1}\\
	&&\gamma^\mu\psi_\mu(p)=0,\label{Eq:eom2}\\
	&&p^\mu\psi_\mu(p)=0.\label{Eq:eom3}
\end{eqnarray}
In the following discussion, we will use these equations to simplify the amplitudes whenever possible.

First, let us consider diagrams A and B. 
For a generic mass spectrum, the amplitudes are given by
\begin{eqnarray}
	i\hat{\cal M}_A &=& \frac{1}{2}C_{t SS}\int\frac{d^4q}{(2\pi)^4}\overline u_\mu(p)P_L\frac{\cancel{q}\g^\mu\cancel{q}\g^\nu\cancel{q}}{D_0D_1D_2}v_\nu(k),\\
	i\hat{\cal M}_B &=& -\frac{1}{2}C_{t FF}\mu\int\frac{d^4q}{(2\pi)^4}\overline u_\mu(p)P_L
	\left[
		2\frac{\cancel{q}\g^\mu\cancel{q}\g^\nu\cancel{q}}{D_0D_1D_2}\right.\nonumber\\
	&&\left.
		+\frac{\cancel{q}\g^\mu(\cancel{p}_1+\cancel{p}_2)\g^\nu\cancel{q}}{D_0D_1D_2}
	\right]
	v_\nu(k),
\end{eqnarray}
where we have omitted the factors arising from changing the basis of Higgs and Higgsinos into neutralinos and charginos, and the propagator in the denominators are defined as $D_0=q^2-m_0^2$ and $D_i=(q+p_i)^2-m_i^2~(i=1,2)$.
The gravitino wave functions are denoted as $u_\mu$ and $v_\nu$ which satisfy the Majorana condition $u_\mu=v_\mu^c$.
The momenta in the propagators are taken to be $p_1=p$ and $p_2=-k$, while $m_0$ and $m_1(=m_2)$ are the Higgsino and Higgs boson masses for diagram A, and are the Higgs and Higgsino masses for diagram B, respectively.
Then, the resultant amplitudes are given by
\begin{eqnarray}
	i\hat{\cal M}_A &=& \frac{iC_{t SS}}{32\pi^2}
	\left[
		C_{001}^A ~({\rm I+II+III})+C_{002}^A ~({\rm I+II+IV})
	\right.\nonumber\\
		&&+C_{111}^A ~{\rm V}+C_{222}^A ~{\rm VI}
	\nonumber\\
		&&\left.
		+C_{112}^A ~({\rm VII+VIII+IX})+C_{122}^A ~({\rm X+XI+XII})
	\right]\nonumber\\
	&=&\frac{iC_{t SS}}{32\pi^2}
	\left[
		C_{001}^A ~{\rm III}+C_{002}^A ~{\rm IV}+C_{112}^A ~{\rm IX}+C_{122}^A ~{\rm X}
	\right],\nonumber\\
	&&\\
	i\hat{\cal M}_B &=& -2i\hat{\cal M}_{A\leftrightarrow B}\nonumber\\
	&& -\frac{iC_{t FF}\mu}{32\pi^2}
	\left[
		C_{00}^B ~({\rm III+IV})+C_{11}^B ~({\rm V+VIII})
	\right.\nonumber\\
	&&\left.
	   +C_{22}^B ~({\rm XI+VI})+C_{12}^B ~({\rm VII+IX+X+XII})
	\right]\nonumber\\
	&=&-\frac{iC_{t FF}\mu}{32\pi^2}
	\left[
		(C_{00}^B+2C_{001}^B) ~{\rm III}+(C_{00}^B+2C_{002}^B) ~{\rm IV}
	\right.\nonumber\\
	&&\left.
	+(C_{12}^B+2C_{112}^B) ~{\rm IX}+(C_{12}^B+2C_{122}^B) ~{\rm X}
	\right],
\end{eqnarray}
where $C_{ij}^I$ and $C_{ijk}^I$ are the Passarino-Veltman functions \cite{Passarino:1978jh} for the diagram $I=A, B$, which are defined as
\begin{eqnarray}
	C_\alpha;C_{\alpha\beta};C_{\alpha\beta\gamma}
	&=&
	\int\frac{d^4q}{i\pi^2}\frac{q_\alpha;q_\alpha q_\beta;q_\alpha q_\beta q_\gamma}{D_0 D_1 D_2}.
\end{eqnarray}
These functions may in general be written as
\begin{eqnarray}
	C_\alpha &=& p_{1\alpha}C_1+p_{2\alpha}C_2,\\
	C_{\alpha\beta} &=& g_{\alpha\beta}C_{00}+p_{1\alpha}p_{1\beta}C_{11}+p_{2\alpha}p_{2\beta}C_{22}\nonumber\\
	&&+\{p_{1\alpha}p_{2\beta}+p_{2\alpha}p_{1\beta}\}C_{12},\\
	C_{\alpha\beta\gamma} &=& \sum_{i=1,2}\{g_{\alpha\beta}p_{i\gamma}+g_{\beta\gamma}p_{i\alpha}+g_{\gamma\alpha}p_{i\beta}\}C_{00i}\nonumber\\
	&&+
	\{p_{1\alpha}p_{1\beta}p_{2\gamma}+p_{1\alpha}p_{2\beta}p_{1\gamma}+p_{2\alpha}p_{1\beta}p_{1\gamma}\}C_{112}\nonumber\\
	&&+
	\{p_{2\alpha}p_{2\beta}p_{1\gamma}+p_{2\alpha}p_{1\beta}p_{2\gamma}+p_{1\alpha}p_{2\beta}p_{2\gamma}\}C_{122}.\nonumber\\
	&&
\end{eqnarray}
The terms denoted by I, II, III, ... are the amplitudes simplified by the EOM and the relevant relations, and are given by
\begin{enumerate}[I.]
	\item $\overline u_\mu P_L \g^\a\g^\mu\g_\a\g^\nu\cancel{p}_i v_\nu=0$,
	\item $\overline u_\mu P_L \cancel{p}_i\g^\mu\g^\b\g^\nu\g_\b v_\nu=0$,
	\item $\overline u_\mu P_L \g^\a\g^\mu\cancel{p}_1\g^\nu\g_\a u_\nu=4m_{3/2}g^{\mu\nu}\overline u_\mu P_R v_\nu$,
	\item $\overline u_\mu P_L \g^\a\g^\mu\cancel{p}_2\g^\nu\g_\a u_\nu=4m_{3/2}g^{\mu\nu}\overline u_\mu P_L v_\nu$,
	\item $\overline v_\mu P_L \cancel{p}_1\g^\mu\cancel{p}_1\g^\nu\cancel{p}_1v_\nu=0$,
	\item $\overline u_\mu P_L \cancel{p}_2\g^\mu\cancel{p}_2\g^\nu\cancel{p}_2v_\nu=0$,
	\item $\overline u_\mu P_L \cancel{p}_1\g^\mu\cancel{p}_1\g^\nu\cancel{p}_2v_\nu=0$,
	\item $\overline u_\mu P_L \cancel{p}_1\g^\mu\cancel{p}_2\g^\nu\cancel{p}_2v_\nu=0$,
	\item $\overline u_\mu P_L \cancel{p}_2\g^\mu\cancel{p}_1\g^\nu\cancel{p}_1v_\nu=-4m_{3/2}k^\mu p^\nu\overline u_\mu P_R v_\nu$,
	\item $\overline u_\mu P_L \cancel{p}_2\g^\mu\cancel{p}_2\g^\nu\cancel{p}_1v_\nu=-4m_{3/2}k^\mu p^\nu\overline u_\mu P_L v_\nu$,
	\item $\overline u_\mu P_L \cancel{p}_2\g^\mu\cancel{p}_1\g^\nu\cancel{p}_2v_\nu=0$,
	\item $\overline u_\mu P_L \cancel{p}_1\g^\mu\cancel{p}_2\g^\nu\cancel{p}_2v_\nu=0$,
\end{enumerate}
where the momentum for $\overline u_\mu$ and $v_\nu$ is assigned to be $\overline u_\mu(p)$ and $v_\nu(k)$.

For diagram C, we have
\begin{eqnarray}
	i\hat{\cal M}_C &=& \frac{1}{4}C_{t SS}c_H \mu\int\frac{d^4q}{(2\pi)^4}\overline u_\mu(p)P_L\frac{\Gamma^{\mu\sigma}}{D_0D_1}v_\sigma(k),
\end{eqnarray}
where $c_H$ represents a constant factor coming from the mixing of Higgs bosons, and the vertex factor is defined as
\begin{eqnarray}
	\Gamma^{\mu\sigma} &=& -\frac{1}{2}[\gamma^\mu,\gamma^\sigma].
\end{eqnarray}
Using Eq. (\ref{Eq:eom2}), we obtain
\begin{eqnarray}
	i\hat{\cal M}_C&=&\frac{1}{8}\frac{i}{16\pi^2}C_{t SS}\mu c_H B_0 g^{\mu\nu}\overline u_\mu v_\nu,
\end{eqnarray}
where $B_0$ is the two-point scalar function defined as
\begin{eqnarray}
	B_0 &\equiv& \int\frac{d^4q}{i\pi^2}\frac{1}{D_0D_1}.
\end{eqnarray}
The coupling $c_H$ is found by recalling the relation between $H_{u,d}$ and their mass eigenstates\footnote{We denote the lightest neutral Higgs $h^0$, heavier neutral Higgs $H^0$, CP-odd Higgs $A$, charged Higgs $H^\pm$, neutral Goldstone boson $G^0$, and charged Goldstone boson $G^\pm$.} given as
\begin{eqnarray}
	H_u^0 &=& \frac{1}{\sqrt{2}} (c_\alpha h^0+s_\alpha H^0+is_\beta G^0+ic_\beta A),\\
	H_d^0 &=& \frac{1}{\sqrt{2}} (-s_\alpha h^0+c_\alpha H^0-ic_\beta G^0+is_\beta A),\\
	H_u^+ &=& s_\beta G^++c_\beta H^+,\\
	H_d^{-*} &=& -c_\beta G^++s_\beta H^+
\end{eqnarray}
with $s_x\equiv \sin x$ and $c_x\equiv \cos x$.
We find
\begin{eqnarray}
	c_HB_0 &=& -\frac{1}{2}s_\alpha c_\alpha B_0(h^0);+\frac{1}{2}s_\alpha c_\alpha B_0(H^0);\nonumber\\
	&&+\frac{1}{2}s_\beta c_\beta B_0(G^0);-\frac{1}{2}s_\beta c_\beta B_0(A);\nonumber\\
	&&-s_\beta c_\beta B_0(G^\pm);+s_\beta c_\beta B_0(H^\pm).
\end{eqnarray}
In the $m_A\gg m_Z$ limit and $\tan\beta\simeq 1$, we have $-s_\alpha c_\alpha\sim s_\beta c_\beta \sim 1/2$.
Therefore, in the supersymmetric limit, all of the scalar masses are degenerate in mass at $\mu$, and the sum over all contributions identically vanishes.
In the case that $m_{H^0}\sim m_A\sim m_{H^\pm} \sim \mu \gg m_{h^0}$ and $m_{G^0}=m_{G^\pm}=0$, we have
\begin{eqnarray}
	\sum c_H B_0 &\simeq& \frac{m_{h^0}^2}{M_T^2}\ll 1,
\end{eqnarray}
so the contribution from diagram C turns out to be negligible.

For diagram D, the amplitude is given by
\begin{eqnarray}
	i\hat{\cal M}_D &=& -\frac{1}{8}C_{t FF}\int\frac{d^4q}{(2\pi)^4}\overline u_\mu(p)\Gamma^{\mu\nu\sigma}v_\nu(k)\nonumber\\
	&&\times
	\frac{{\rm tr}[(\cancel{q}+m)(\cancel{q}+\cancel{p}_t+m)\gamma_\sigma P_L]}{D_0 D_1},
\end{eqnarray}
where $p_t$ is the inflaton momentum, and
\begin{eqnarray}
	\Gamma^{\mu\nu\sigma} &=& i\epsilon^{\mu\alpha\nu\sigma}\gamma_\alpha+g^{\mu\nu}\gamma^\sigma\gamma^5.
\end{eqnarray}
After computing the trace in the amplitude, the loop integral is
\begin{eqnarray}
	\int\frac{d^4q}{(2\pi)^4}\frac{2q_\sigma+p_{t,\sigma}}{D_0D_1}
	&=& 
	\frac{i}{16\pi^2}p_{t,\sigma}(2B_1+B_0),
\end{eqnarray}
where $B_1$ is defined as
\begin{eqnarray}
    \int\frac{d^4q}{i\pi^2}\frac{q_\mu}{D_0D_1} &=& p_{1\mu}B_1.
\end{eqnarray}
The spinor piece in the amplitude then gives
\begin{eqnarray}
	\overline u_\mu(p)\Gamma^{\mu\rho\sigma}p_{t,\sigma}v_\rho(k) &=& \overline u_\mu[i\epsilon^{\mu\nu\rho\sigma}\gamma_\nu p_{t,\rho}+g^{\mu\rho}\cancel{p}_t\gamma^5]v_\rho.\nonumber\\
	&&
\end{eqnarray}
The first term may be written as
\begin{eqnarray}
	-2m_{3/2}g^{\mu\rho}\overline u_\mu(p)\gamma^5 v_\rho(k),
\end{eqnarray}
where we have used the EOM.
For the second term we obtain
\begin{eqnarray}
	+2m_{3/2}g^{\mu\rho}\overline u_\mu(p)\gamma^5 v_\rho(k),
\end{eqnarray}
where we have used $\cancel{k}v_\rho(k)=-m_{3/2}v_\rho(k)$ and $\overline u_\mu(p)\cancel{p}=m_{3/2}\overline u_\mu(p)$.
Therefore, the amplitude vanishes for diagram D.

Hence, the total amplitude, whose dominant contributions come from diagrams A and B, can be written as
\begin{eqnarray}
	i\hat{\cal M}_{\rm tot} &=& 2\frac{im_{3/2}}{8\pi^2}
	\left[
		c_{1R}\hat{\cal M}_{1R}+
		c_{1L}\hat{\cal M}_{1L}+
	\right.\nonumber\\
	&&\left.
		c_{2R}\hat{\cal M}_{2R}+
		c_{2L}\hat{\cal M}_{2L}
	\right],
\end{eqnarray}
where the overall factor 2 comes from the exchange of final state gravitinos, and we define
\begin{eqnarray}
	{\rm III} &=& 4m_{3/2}\hat{\cal M}_{1R},\\
	{\rm IV} &=& 4m_{3/2}\hat{\cal M}_{1L},\\
	{\rm IX} &=& 4m_{3/2}\hat{\cal M}_{2R},\\
	{\rm X} &=& 4m_{3/2}\hat{\cal M}_{2L}
\end{eqnarray}
with the coefficients being
\begin{eqnarray}
	c_{1R} &=& C_{t SS}C^A_{001}-\mu C_{t FF}(C^B_{00}+2C^B_{001}),\\
	c_{1L} &=& C_{t SS}C^A_{002}-\mu C_{t FF}(C^B_{00}+2C^B_{002}),\\
	c_{2R} &=& C_{t SS}C^A_{112}-\mu C_{t FF}(C^B_{12}+2C^B_{112}),\\
	c_{2L} &=& C_{t SS}C^A_{122}-\mu C_{t FF}(C^B_{12}+2C^B_{122}).
\end{eqnarray}
By denoting $F_{i,j}\equiv {\cal M}_i^*{\cal M}_j$ with $i=1R,1L,2R,2L$, we have
\begin{eqnarray}
	F_{1R,1R} &=& \frac{1}{3}M_T^2\tau,\\
	F_{1L,1L} &=& F_{1R,1R},\\
	F_{1R,1L} &=& -\frac{1}{9}M_T^2\tau^{-2}(-2+\tau)(8-8\tau+\tau^2),\\
	F_{2R,2R} &=& \frac{2}{9}M_T^6\tau^{-1}(-1+\tau)^2,\\
	F_{2L,2L} &=& F_{2R,2R},\\
	F_{2R,2L} &=& -\frac{2}{9}M_T^6\tau^{-2}(-2+\tau)(-1+\tau)^2,\\
	F_{1R,2R} &=& -\frac{1}{9}M_T^4\tau^{-1}(-2+\tau)(-1+\tau),\\
	F_{1L,2L} &=& F_{1R,2R},\\
	F_{1R,2L} &=& \frac{1}{9}M_T^4\tau^{-2}(-8+16\tau-9\tau^2+\tau^3),\\
	F_{1L,2R} &=& F_{1R,2L},
\end{eqnarray}
where $\tau\equiv 4m_{3/2}^2/M_T^2$.

Next, let us express all Higgs and Higgsinos as mass eigenstates.\footnote{We denote the Higgsino-like chargino as $\widetilde\chi_2^\pm$, and $\widetilde\chi_{3,4}^0$ are the Higgsino-like neutralinos.}
The amplitude for A and B is given as
\begin{eqnarray}
	{\cal M}_I &=& 
	\hat{\cal M}_I(\widetilde\chi_2^\pm,H^\pm)
	+\hat{\cal M}_I(\widetilde\chi_2^\pm,G^\pm)\nonumber\\
	&&
	+\frac{1}{4}
	\sum_{i=3,4}
	\left[
		\hat{\cal M}_I(\widetilde\chi_i^0,h^0)
		+\hat{\cal M}_I(\widetilde\chi_i^0,H^0)
	\right.\nonumber\\
	&&\left.
		+\hat{\cal M}_I(\widetilde\chi_i^0,G^0)
		+\hat{\cal M}_I(\widetilde\chi_i^0,A)
	\right]
\end{eqnarray}
for $I=A,B$, and thus we have,
\begin{eqnarray}
	{\cal M}_{\rm A+B} &=& 
	\hat{\cal M}_{\rm A+B}(\widetilde\chi_2^\pm,H^\pm)
	+\hat{\cal M}_{\rm A+B}(\widetilde\chi_2^\pm,G^\pm)\nonumber\\
	&&
	+\frac{1}{4}
	\sum_{i=3,4}
	\left[
		\hat{\cal M}_{\rm A+B}(\widetilde\chi_i^0,h^0)
		+\hat{\cal M}_{\rm A+B}(\widetilde\chi_i^0,H^0)
	\right.\nonumber\\
    &&\left.
		+\hat{\cal M}_{\rm A+B}(\widetilde\chi_i^0,G^0)
		+\hat{\cal M}_{\rm A+B}(\widetilde\chi_i^0,A)
	\right].
\end{eqnarray}
As a good approximation for high-scale supersymmetry, let us suppose that $m_{\widetilde \chi_{2}^\pm}\sim m_{\chi_{3,4}^0}\sim m_{H^\pm}\sim m_{H^0}\sim m_A\sim \mu\gg m_{h^0}$.
Then, we may write the total amplitude as
\begin{eqnarray}
	{\cal M}_{\rm tot} &\simeq& 2{\cal M}_{\rm tot}^h+2{\cal M}_{\rm tot}^H,
\end{eqnarray}
where ${\cal M}_{\rm tot}^h\equiv\hat{\cal M}_{A+B}(\widetilde\chi^0_i,h^0)=\hat{\cal M}_{A+B}(\widetilde\chi^0_i,G^0)=\hat{\cal M}_{A+B}(\widetilde\chi^0_i,G^\pm)$ and ${\cal M}_{\rm tot}^H\equiv\hat{\cal M}_{A+B}(\widetilde\chi_2^\pm,H^\pm)=\hat{\cal M}_{A+B}(\widetilde \chi^0_i,H^0)=\hat{\cal M}_{A+B}(\widetilde \chi^0_i,A)$.

We can further simplify the amplitude by taking the limit $m_{3/2}\ll M_T \ll \mu$.
In this limit, the leading amplitudes among $F_{i,j}$ are $F_{1R,1L}, F_{2R,2L}$, and $F_{1R,2L}$ for $\tau \ll 1$, namely,
\begin{eqnarray}
	F_{1R,1L} &\simeq& \frac{16}{9}M_T^2\tau^{-2},\\
	F_{2R,2L} &\simeq& \frac{4}{9}M_T^6\tau^{-2},\\
	F_{1R,2L} &\simeq& -\frac{8}{9}M_T^4\tau^{-2}.
\end{eqnarray}
Besides, as we can see from the dimensionality, $C_{00}$ and $C_{00i}$ in $c_{1R}$ and $c_{1L}$ remain constant (or logarithmic of the involved mass scales, i.e., $\mu$) as it is dimensionless, while $C_{12}^B$ in $c_{2R}$ and $c_{2L}$ scales as $1/\mu^2$.
Indeed, we obtain analytic expressions of those functions in this limit:
\begin{eqnarray}
    C_{001}^{A,h} &\simeq& -\frac{1}{9}+\frac{1}{12}\ln\frac{\mu^2}{\mu^2_{\rm ren}},\\
    C_{001}^{A,H} &\simeq& \frac{1}{24}+\frac{1}{12}\ln\frac{\mu^2}{\mu^2_{\rm ren}},\\
	C_{00}^{B,h} &\simeq&
	-\frac{1}{4}\ln\frac{\mu^2}{\mu^2_{\rm ren}},\\
	C_{00}^{B,H} &\simeq& -\frac{1}{8}-\frac{1}{4}\ln\frac{\mu^2}{\mu^2_{\rm ren}},\\
	C_{001}^{B,h} &\simeq& \frac{1}{72}+\frac{1}{12}\ln\frac{\mu^2}{\mu^2_{\rm ren}},\\
	C_{001}^{B,H} &\simeq& \frac{1}{24}+\frac{1}{12}\ln\frac{\mu^2}{\mu^2_{\rm ren}},
\end{eqnarray}
and $C_{001}^{I,h(H)}=C_{002}^{I,h(H)}$ for $I=A,B$.
Therefore, the dominant contribution in the decay width is the term involving ${\rm Re}(c_{1R}^*c_{1L}) F_{1R,1L}$ in the squared amplitude.
Then, we obtain
\begin{eqnarray}
	\Gamma_{\phi\to3/2}^{\rm loop} &=& \frac{|{\cal M}_{\rm tot}|^2}{2\cdot16\pi M_T}
	\simeq
	\frac{{\rm Re}(\bar c_{1R}^* \bar c_{1L})}{3^24^4\pi^5}\frac{M_T^5}{m_{3/2}^2}\nonumber\\
	&\simeq&
	\frac{2}{3^34^5\pi^5}
	\left(
		\frac{1}{4}-\ln\frac{\mu^2}{\mu^2_{\rm ren}}
	\right)^2
	\frac{\mu^4 M_T^5}{m_{3/2}^2M_P^6},
\end{eqnarray}
where $\bar c_i\equiv c_i^h+c_i^H$ for $i=1R,1L,2R,2L$, and we have used
\begin{eqnarray}
	\bar c_{1R} \simeq \bar c_{1L} 
	&\simeq&
	-\frac{C_{t SS}}{2}
	\left[
		2C_{00}^{B,h}+2C_{00}^{B,H}
	\right]\nonumber\\
	&\simeq&
	-\frac{1}{2}\sqrt{\frac{2}{3}}\mu^2
	\left(
		\frac{1}{4}-\ln\frac{\mu^2}{\mu^2_{\rm ren}}
	\right)
\end{eqnarray}
with the relation $C_{tSS}=2\mu C_{tFF}$.

\vspace{-.5cm}
\bibliographystyle{apsrev4-1}

\end{document}